\PassOptionsToPackage{pdfpagelabels=false}{hyperref}
\documentclass[fleqn,usenatbib]{mnras}
\usepackage{xcolor, graphicx}
\usepackage{amsmath}
\usepackage{amssymb}
\usepackage{soul}
\usepackage{txfonts}
\usepackage{journal_names}
\usepackage{caption}
\usepackage{todonotes}
\usepackage[export]{adjustbox}
\usepackage{longtable}

\newcommand{\Msun}{\rm M_\odot}
\newcommand{\vlos}{V_{\rm los}}
\newcommand{\kms}{\mbox{km s$^{-1}$}}

\newcommand{\SBunit}{\mbox{mag arcsec$^{-2}$}}

\def\Sersic/{{S\'ersic}}

\title[GCs in the stellar stream surrounding the MW analog NGC~5907]
{Globular clusters in the stellar stream surrounding the Milky Way analog NGC~5907}
\author[Alabi~et~al.~ ]
{Adebusola B. Alabi$^{1}$\thanks{Email: aalabi@ucsc.edu}, Duncan A. Forbes$^{2}$,  Aaron J. Romanowsky$^{1,3}$, Jean P. Brodie$^{1}$\\
\\
$^{1}$ University of California Observatories, 1156 High Street, Santa Cruz, CA 95064, USA\\
$^{2}$ Centre for Astrophysics \& Supercomputing, Swinburne University, Hawthorn VIC 3122, Australia\\
$^{3}$ Department of Physics and Astronomy, San Jos\'e State University, San Jose, CA 95192, USA\\
}

\begin{document}

\date{Accepted XXX. Received YYY; in original form ZZZ}
\pagerange{\pageref{firstpage}--\pageref{lastpage}} \pubyear{2019}
\maketitle

\label{firstpage}
\begin{abstract}
We study the globular clusters (GCs) in the spiral galaxy NGC~5907 well-known for its spectacular stellar stream -- to better understand its origin. Using wide-field  Subaru/Suprime-Cam $gri$ images and deep Keck/DEIMOS multi-object spectroscopy, we identify and obtain the kinematics of several GCs superimposed on the stellar stream and the galaxy disk. We estimate the total number of globular clusters in NGC~5907 to be $154\pm44$, with a specific frequency of $0.73\pm0.21$. Our analysis also reveals a significant, new population of young star cluster candidates found mostly along the outskirts of the stellar disk. Using the properties of the stream GCs, we estimate that the disrupted galaxy has a stellar mass similar to the Sagittarius dwarf galaxy accreted by the Milky Way, i.e. $\sim10^8~\Msun$.
\end{abstract}

\section{Introduction}

In the paradigm of hierarchical galaxy growth over cosmic time, it is 
expected that the signatures of disrupted, and disrupting, satellites 
will be present in galaxy halos \citep{Bullock_2005}. Witnessing the ongoing disruption of a satellite galaxy in the nearby Universe is rare due to the low frequency of occurrence combined with observational challenges given that the surface brightness of the debris declines significantly with time \citep[see][]{Johnston_2001}. Nevertheless a few nearby examples are known \citep[e.g.][]{Malin_1997, Forbes_2003,Mihos_2005,Tal_2009,Duc_2015}, including  notable contributions from amateur telescopes \citep{Delgado_2008, Delgado_2010,Karachentsev_2014}. 

\begin{figure*}
    \includegraphics[width=0.98\textwidth]{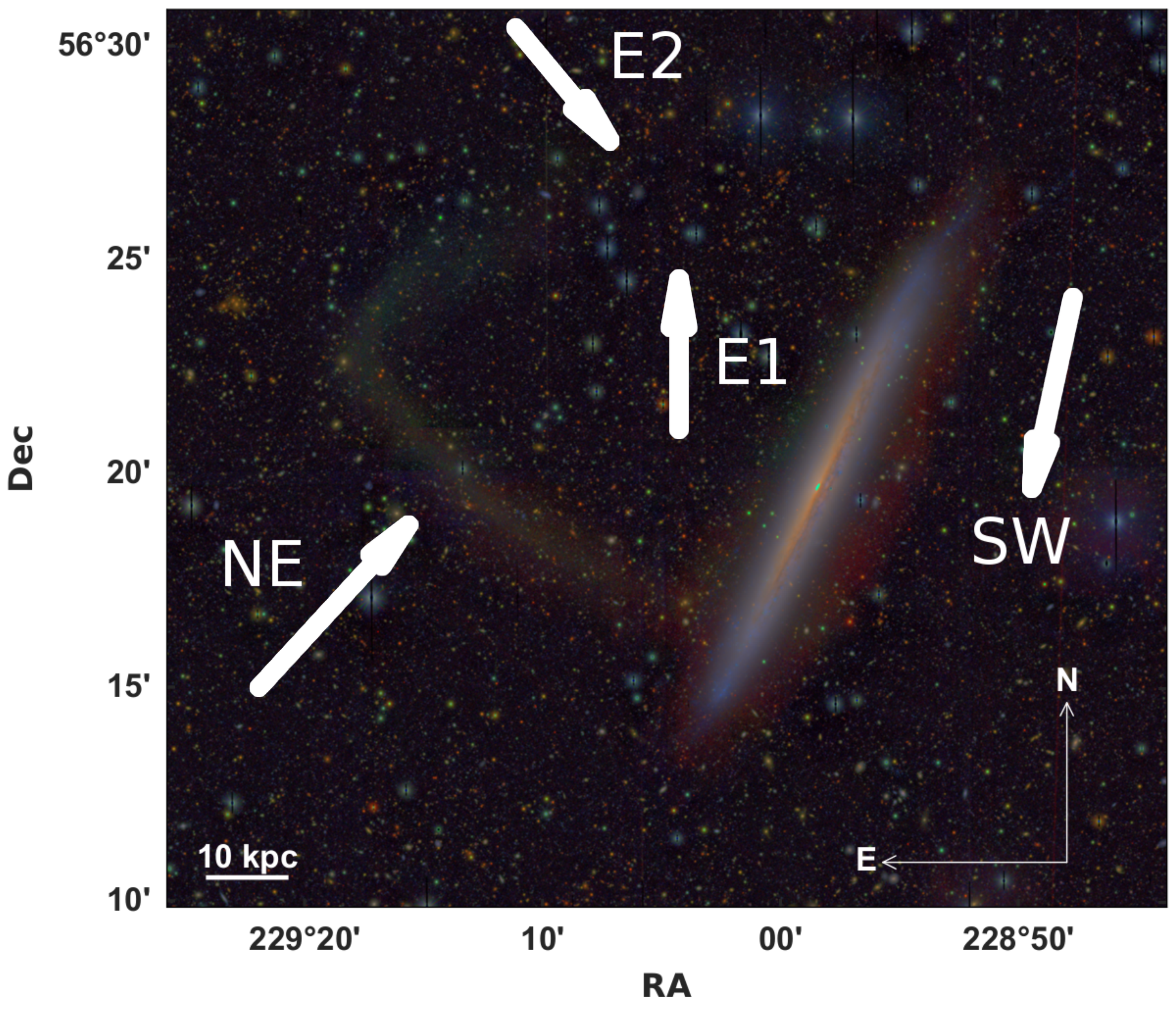}\hspace{0.01\textwidth}\\
	\caption{\label{fig:gal_img} Colorized composite image of NGC~5907 from Subaru/Suprime-Cam $g,r$, and $i$ filters. The image covers $115 \times 100$ kpc and shows the prominent stellar stream in the North-East direction. The loops of the stellar stream have been labeled as referenced in the text and in the literature. The stream structure in our imaging does not reveal the extension in the South-West direction seen by \citet{vanDokkum_2019} nor does it reveal the additional E1 and E2 loops seen in \citet{Delgado_2008}.}
\end{figure*}

One of the most prominent examples is the stream associated with a disrupting satellite around the edge-on Milky Way analog, NGC 5907. It was originally imaged by \citet{Shang_1998} but is perhaps best known from the deep imaging of \citet{Delgado_2008}. This latter work shows several loops around the galaxy. \citet{Laine_2016} focused on the brightest part of the stream which lies to the NE of NGC 5907, extending out to $\sim 65$~kpc as shown in Figure~\ref{fig:gal_img}. 
Most recently deep imaging using the Dragonfly telephoto array has revealed a different view with a single long stream stretching from the NE to the SW of the galaxy \citep{vanDokkum_2019}. It is not clear why the Dragonfly imaging differs from that of \citet{Delgado_2008}.

Similarly, details of the progenitor satellite responsible for the stellar stream(s) in NGC~5907 are still poorly defined, with the inferred mass of the disrupted satellite galaxy varying in the literature. The study of \citet{Laine_2016}, based on stellar populations in the NE stream, concluded that the stellar mass of the disrupting satellite was $\sim10^{10}~\Msun$, i.e., a 1:10 minor merger. This is considerably larger than the mass of $\sim$10$^{8}~\Msun$ in the semi-analytic modeling of \citet{Johnston_2001} and the \textit{N}-body models of \citet{Delgado_2008} and \citet{vanDokkum_2019}. \citet{Wang_2012} also used a 1:3 gas-rich major merger to model the stellar loops in NGC~5907. Thus it is currently uncertain whether the satellite debris results from a major, minor, or intermediate mass merger. 


To further constrain the mass of the progenitor satellite, line-of-sight velocities along the stellar stream(s) are therefore required \citep[e.g.][]{Lynden_1995, Dinescu_1999}. Given the faint surface brightness of even the brightest loop ($\langle \mu_g \rangle \sim 27.6$ mag arcsec$^{-2}$), this is extremely challenging. An alternative is to use globular clusters (GCs) in the stream as proxies. Such an approach was successfully applied by \citet{Foster_2014} to probe the nature of a 1:50 stellar mass ratio merger in the Umbrella galaxy NGC~4651. Similarly, the outer halo GCs in the Galaxy and M31 have been shown to trace the remnants of disrupted satellites \citep{Mackey_2010, Massari_2019}.  
For a satellite with stellar mass 10$^8$ to 10$^{10}$ M$_{\odot}$, the number of associated GCs could be as low as $1$ or as high as $100$ \citep{Forbes_2018}. 

Identifying suitable GC candidates in spiral galaxies is challenging. This is mostly due to issues such as dust, disk, and bulge obscurations, and their numerically poor GC systems. For example, \citet{Harris_1988} reported a non-detection of GCs in NGC~5907 from their ground-based observations. \citet{KP_1999}, using \textit{HST}/WFPC2, detected 25 GCs near the disk of NGC~5907, although their imaging did not extend into the stream area. After a correction for spatial incompleteness based on the Milky Way's GC system, they estimated a total GC system of 170$^{+47}_{-72}$. 

In this study, we present wide-field Subaru imaging of NGC~5907 and Keck spectroscopy of several GC candidates. We assume a distance of $17$~Mpc to NGC~5907, as adopted by \citet{Laine_2016} and summarize salient properties of NGC~5907 in Table~\ref{tab:tab_gal}. Many of its properties are similar to those of the Milky Way.
The paper is organized as follows: Section~\ref{partI} describes the imaging data, source detection, photometry, and spectroscopic data. In Section~\ref{partII}, we present our results including the catalog of GC candidates, young star cluster candidates, and stream GCs. We conclude by discussing the implications of our results in Section~\ref{partIII}.

\begin{table}
\begin{tabular}{@{}l c c}
\hline
Property & Value & Reference \\
\hline
\hline
Hubble type & Sc & NED\\
$M_V$ & $-20.81$ & NED\\
Stellar mass   & $1\times10^{11}\ \Msun$    &  \citet{Posti_2019} \\
Total gas mass & $8.6\times10^{9}\ \Msun$     &  \citet{Just_2006}  \\
Disc scale length             & $5.5$~kpc                  &  \citet{Just_2006}  \\
Disc cutoff radius          & $19.7$~kpc                 &  \citet{Just_2006}  \\
Distance       & $17$~Mpc                   &  \citet{Tully_2016} \\
$V_{\rm circ}$   & $226.7$ \kms               &  \citet{Makarov_2014} \\
$V_{\rm sys}$     & $667$ \kms                 &  \citet{Tully_2016} \\
PA             & $155^{\circ}$              &  \citet{Jarrett_2003} \\
Inclination            & $88^{\circ}$               &  \citet{Just_2006} \\
\hline
\end{tabular}
\caption{Properties of NGC~5907}
\label{tab:tab_gal}
\end{table}

\section{DATA}
\label{partI}
\subsection{Observations}
\label{obs}
The imaging used for this work was obtained using the wide-field Suprime-Cam imager on the 8.2m Subaru telescope with the SDSS \textit{gri} filters under photometric conditions ($0''.6 - 0''.9$ seeing). Details of the observations and data reduction are explained in \citet{Laine_2016}. The imager has a total field-of-view of $34' \times 27'$ with a pixel scale of $0''.2$ per pixel. 

In Figure~\ref{fig:gal_img} we show our color composite image of NGC~5907. It clearly shows the NE stellar stream but it is lacking the extension to the SW seen by \citet{vanDokkum_2019} and the E1 and E2 loops of \citet{Delgado_2008}. The surface brightnesses along the faintest section of the stream (E2 in Figure~\ref{fig:gal_img}) are $28.5$, $28.0$, and $27.8$~\SBunit~in our $g$, $r$, and $i$ imaging, respectively. 

\subsection{Image preparation}

For optimal point-source detection, the galaxy light and other large scale structures need to be removed from the raw \textit{gri} images. This is done by subtracting a model of the galaxy  bulge and disk, obtained with \texttt{GALFIT} \citep{Peng_2002} from the raw images before applying a median filter to remove large scale residual features with the ring median filter algorithm described in \citet{Secker_1995}. This method efficiently produces a galaxy image with minimal residuals (down to the galaxy center) suitable for point-source detection. We subtract the median-filtered image\footnote{median-filtered images were created with the \texttt{RMEDIAN} task in \texttt{IRAF} using inner and outer radii of $5$ and $9$ times the FWHM of point sources in the images} from the galaxy-subtracted images and add back the background counts for accurate photon statistics. We call the output of this process our ``cleaned'' images.

\subsection{Source Detection}

Source detection was performed on our ``cleaned'' \textit{gri} images with \textsc{SExtractor} \citep{Bertin_1996}. We detect and perform aperture photometry on $14692$ sources common to our \textit{gri} images. Extended sources (i.e. background galaxies) are removed from our catalog using the following criteria and fine-tuning the steps to accommodate as many as possible of the previously reported GC candidates detected in \textit{HST} imaging \citep{KP_1999}:

\begin{itemize}
	\item \textsc{SExtractor} FLAG parameter $> 0$,
	\item CLASS\_STAR parameter $< 0.3$, and
	\item all FWHM $> 8$ pixels,
\end{itemize}

\noindent
This process gives 5947 point-sources remaining. 

We convert our instrumental magnitudes (\textsc{MAG\_AUTO} from \textsc{SExtractor}) to the SDSS photometric system using bright point-sources ($17 < g < 21$) in common with our catalog. Finally, Galactic extinction corrections in the direction of NGC~5907 from the dust reddening maps of \citet{Schlafly_2011} are applied to our point-source catalog to obtain extinction-corrected magnitudes in each bands. The corrections applied are $A_g=0.035, A_r=0.024, $ and $A_i=0.018$. No internal reddening correction is applied.

\begin{figure}
    \includegraphics[width=0.48\textwidth]{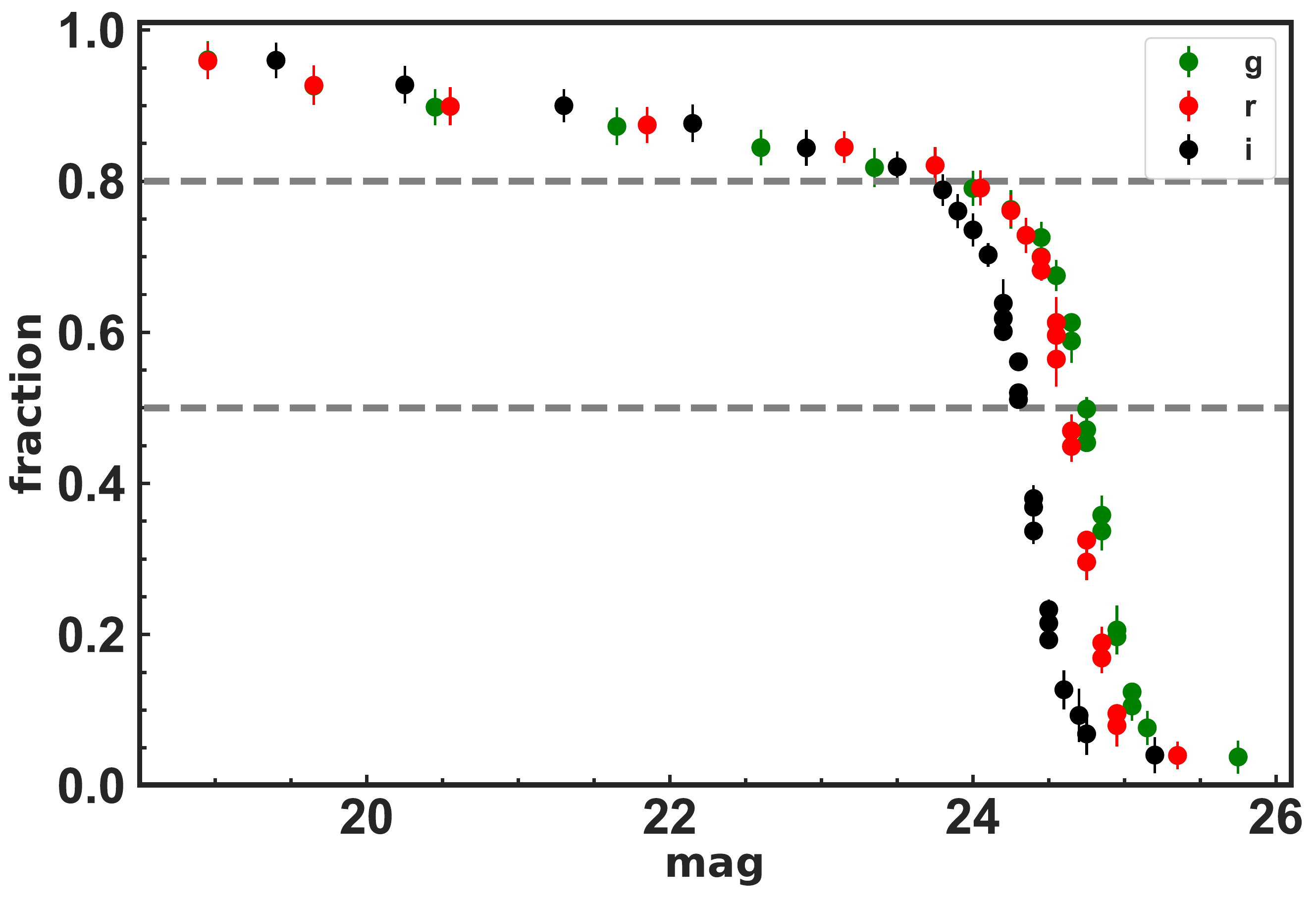}\hspace{0.01\textwidth}\\
	\caption{\label{fig:comp_fig} Detection completeness from artificial star experiment in the $gri$ bands. The dashed lines show the 50\% and 80\% completeness levels.}
\end{figure}

\subsection{Photometric Completeness and Magnitude Limits}
\label{completeness}
To understand how well we detect point-sources as a function of magnitude, we artificially added 600 point-sources to our ``cleaned'' images, with magnitudes uniformly spread over the range $18 < {\rm mag} <26.5$. We then performed source detection as described above, noting the fraction of point-sources recovered. We repeated this experiment 60 times per filter (adding a total of 36000 artificial point-sources) and show the recovered fraction as a function of magnitude, highlighting the 50\% and 80\% completeness levels, in Figure~\ref{fig:comp_fig}. Due to the high variability of the residual background in the disc region after subtracting off the galaxy light, the recovered fraction is typically less than 100\% in the bright magnitude range, i.e., $20 < mag < 23$. Our photometry is 50\% complete at $g=24.8$, $r=24.6$, and $i=24.3$, respectively.

We use the 80\% completeness magnitude in the $i$-band (i.e., $i=23.8$), where detection starts to fall off sharply, to define the faint limit of our photometry. Likewise, we define a bright magnitude limit based on the magnitude of the brightest GC in the Galaxy, $M_i=-11.0$ (see a similar application  in \citealt{Pota_2015}), using a distance modulus $m-M=31.15$. Applying this bright magnitude limit to our catalog ensures that all point-sources brighter than the most luminous GCs in the MW are excluded from subsequent analysis. This brings the number of point-sources to $3798$. At this stage, our point-source catalog includes $11$ objects out of the $25$ \textit{HST}-detected GC candidates \citep{KP_1999}. We have also checked the stream region for objects which are brighter than our bright magnitude limit and found none. Such objects could be viable candidates for the nucleus of the disrupted satellite galaxy.

\begin{figure*}
    \includegraphics[width=0.50\textwidth]{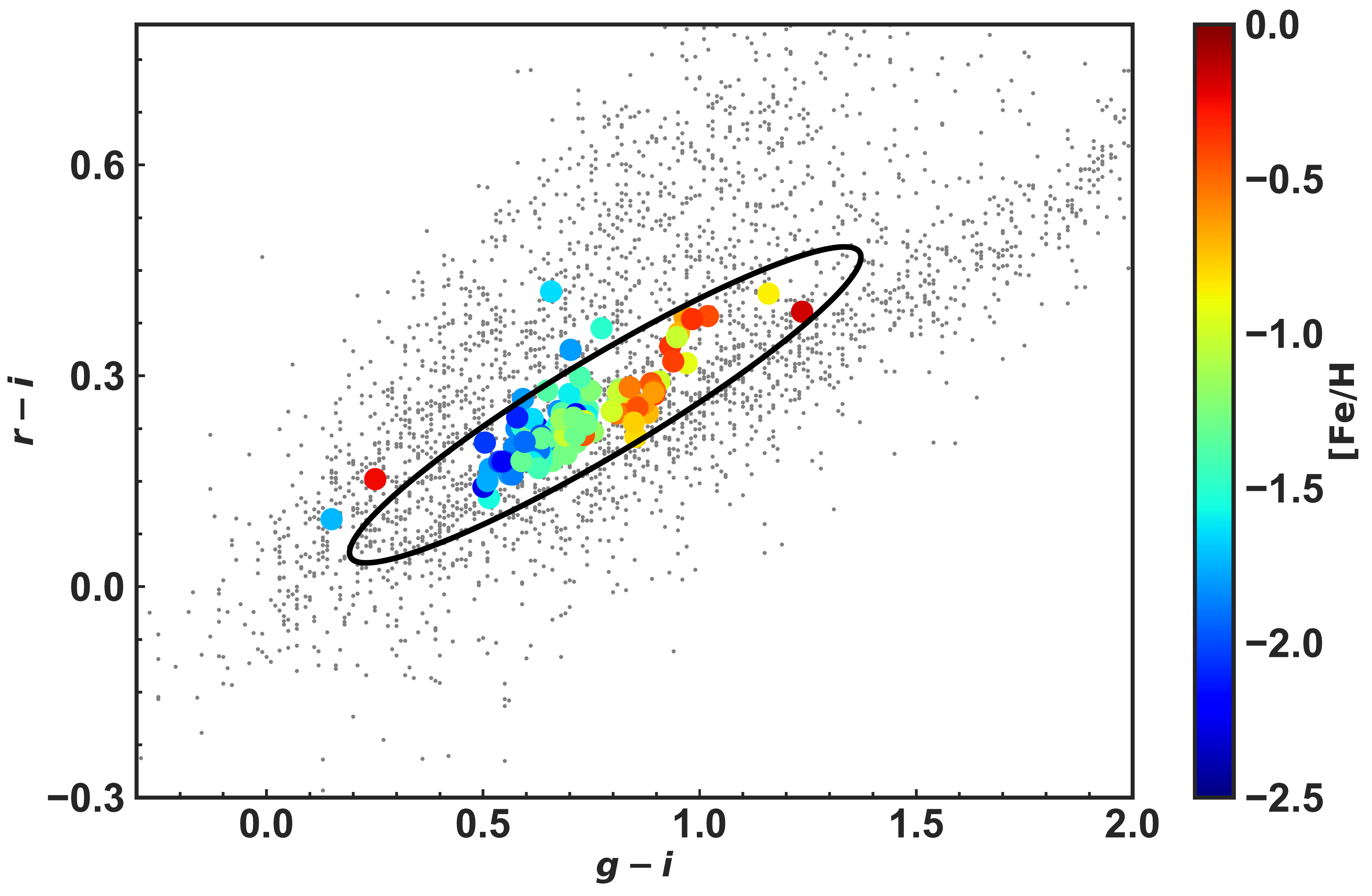}\hspace{0.01\textwidth}
    \includegraphics[width=0.46\textwidth]{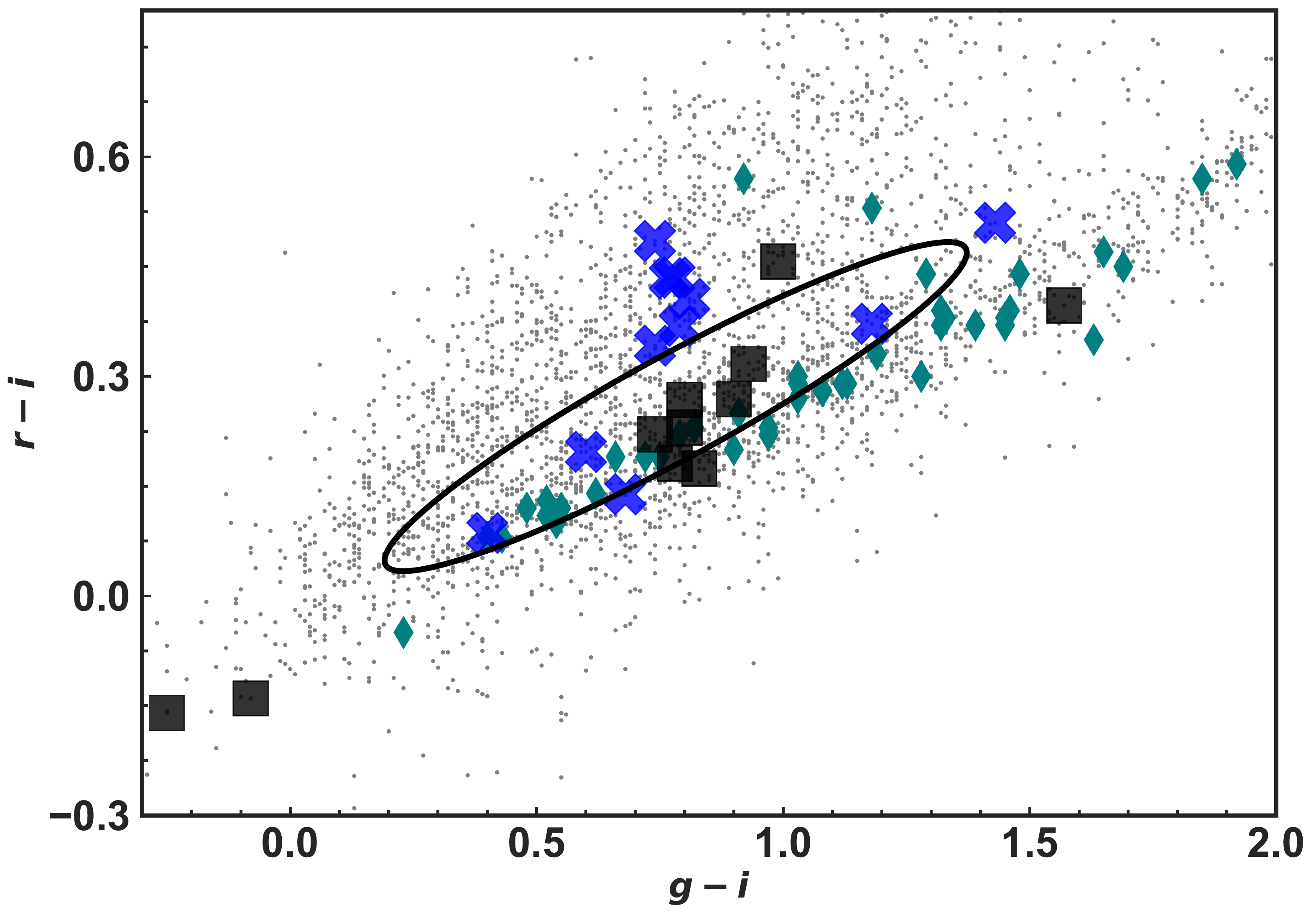}\hspace{0.01\textwidth}\\
	\caption{\label{fig:cmd_fig} Color--color selection of GC candidates around NGC~5907. In the {\it left panel}, the region occupied by Milky Way GCs is used to define a GC selection region (the black ellipse) in color--color space. The Milky Way GCs are shown as filled circles and they are color-coded by their metallicities. The selection region contains $703$ GC candidates. In the {\it right panel}, we show the spectroscopically confirmed GCs around NGC~5907 as black squares, spectroscopic stars as teal diamonds, and the GC candidates that were also detected in the \textit{HST} sample from \citet{KP_1999} as blue crosses. Some of the spectroscopically confirmed point-sources fall outside our GC selection region due to internal reddening and a few are very blue (they may be compact, young star clusters).
	}
\end{figure*}

\subsection{Selection of Globular Cluster Candidates}
\label{color_selection}

Figure~\ref{fig:cmd_fig} shows the $(g-i)$ vs $(r-i)$ color--color plot of all the point-sources detected in NGC~5907. Since GCs are known to occupy distinct regions in color--color space \citep[e.g.,][]{Rhode_2003, Pota_2013, Munoz_2014}, their colors can be used as powerful tools for discriminating them from co-spatial foreground stars and background galaxies. We use the compilation of Milky Way GCs from \citealt[][(2010 edition)]{Harris_1996} to outline the region that GCs occupy in color--color space after converting their $BVR$ photometry to SDSS $gri$ using the transformation equations from \citet{Jester_2005} and correcting for reddening. The transformed and dereddened $(g-i)$ and $(r-i)$~Milky Way GC colors range from $0.27-1.15$ and $0.13-0.41$, respectively, with [Fe/H] spanning from $-2.5$ to $0$. 

Point-sources are selected as GC candidates if they have $i$~band magnitudes ranging from $20.05$ to $23.9$, and $g-i$ and $r-i$ colors similar to the Milky Way GCs. We also visually checked these GC candidates, excluding the obvious background galaxies. Using these criteria, our GC candidate catalog now has $703$ members at an 80\% completeness level in the $i$ band. Out of the $11$ point-sources in common with the \textit{HST}-detected GC sample from \citet{KP_1999}, only $7$ qualify as GC candidates based on our color and magnitude selection criteria. The remaining $4$ \textit{HST}-detected objects have red colors that lie outside our GC color--color selection region. 

While our GC selection here is based on the assumption that the GC system of NGC~5907 is similar to that of the Galaxy, we note that we obtain a similar result if we define our selection region using the GC system of M31 \citep[e.g.,][]{Chen_2016}. We provide a complete summary of all the various classes of point-sources detected in this work in Table~\ref{tab:complete_tab}.

\subsection{Keck Spectroscopy of point-sources around NGC~5907}
\label{spec_GCs}
Using our point-source catalog, we identified target point-sources for follow-up spectroscopy with the DEIMOS spectrograph on the Keck II telescope. 
We obtained three masks; one along with the bright NE loop, one covering the 
E1 and E2 loops of \citet{Delgado_2008}, with the last mask positioned along the disk of NGC~5907. 
In total we placed slits on $359$ point-sources as shown in Figure~\ref{fig:deimos_fig}. 
Data were obtained on the nights of 2013 April 10, 2016 March 10 and 2017 April 27 for a total of $6.5$ hours. We used both the $600$ lines/mm grating (centered on $6700$~\AA) and the $1200$ lines/mm grating (centered on $7800$~\AA). Spectra were reduced using the \textsc{spec2d} pipeline and the same procedure as successfully applied during the SLUGGS survey \citep[e.g.,][]{Pota_2013, Forbes_2017}. 
We measured line-of-sight velocities ($\vlos$) from the CaT and H$\alpha$ spectral features with \textsc{FXCOR} in \textsc{IRAF} using stellar templates from the  MILES spectral library \citep{Vazdekis_2016}.

\begin{figure}
   \includegraphics[width=.48\textwidth]{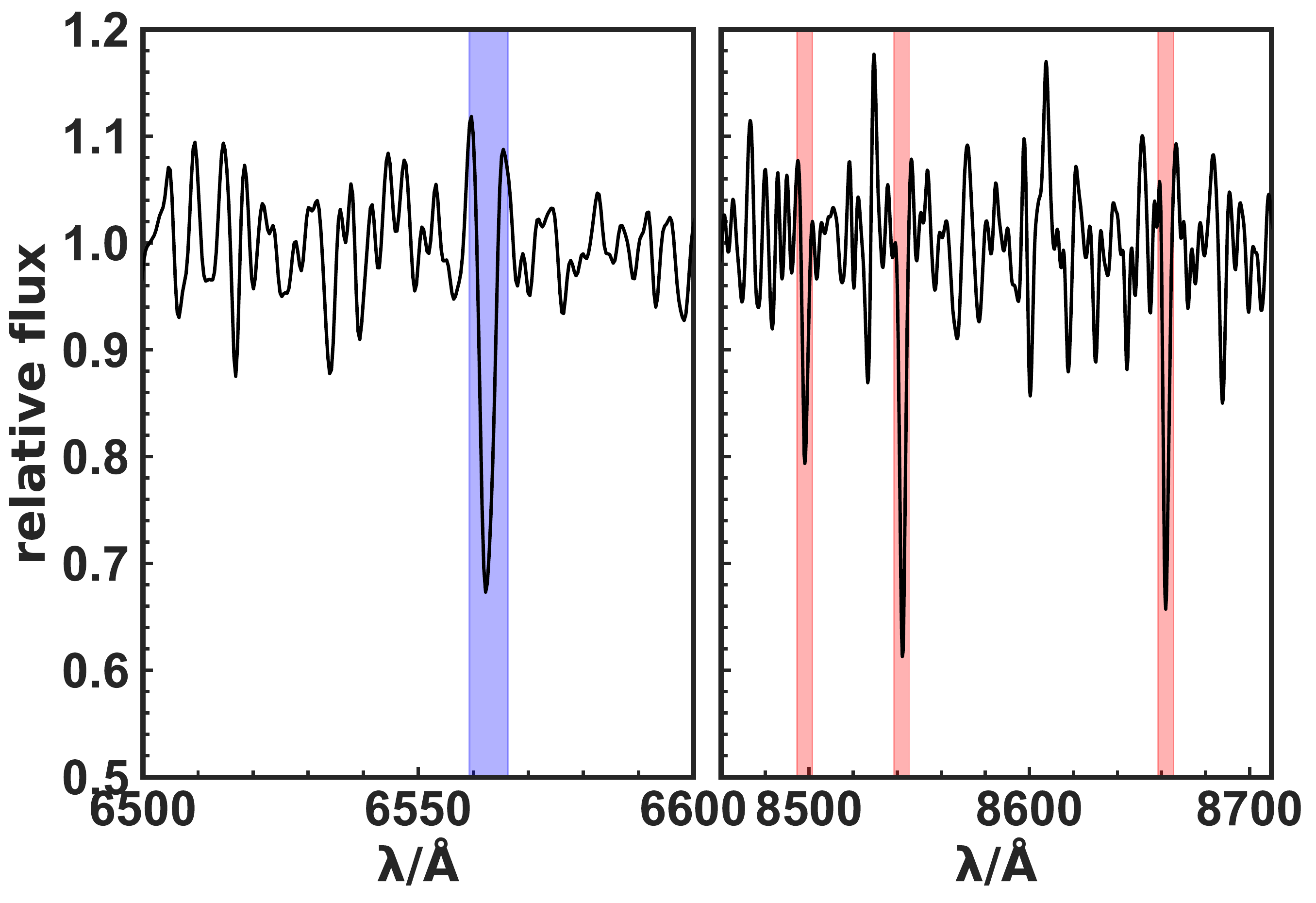}
   \caption{Stacked spectra  (S/N$\sim12$) of the confirmed globular clusters in NGC~5907 highlighting the H$\alpha$ (blue band) and Calcium triplet (red bands) spectral regions.} 
   \label{fig:stacked_spec}
\end{figure}

\begin{figure*}
    \includegraphics[width=0.98\textwidth]{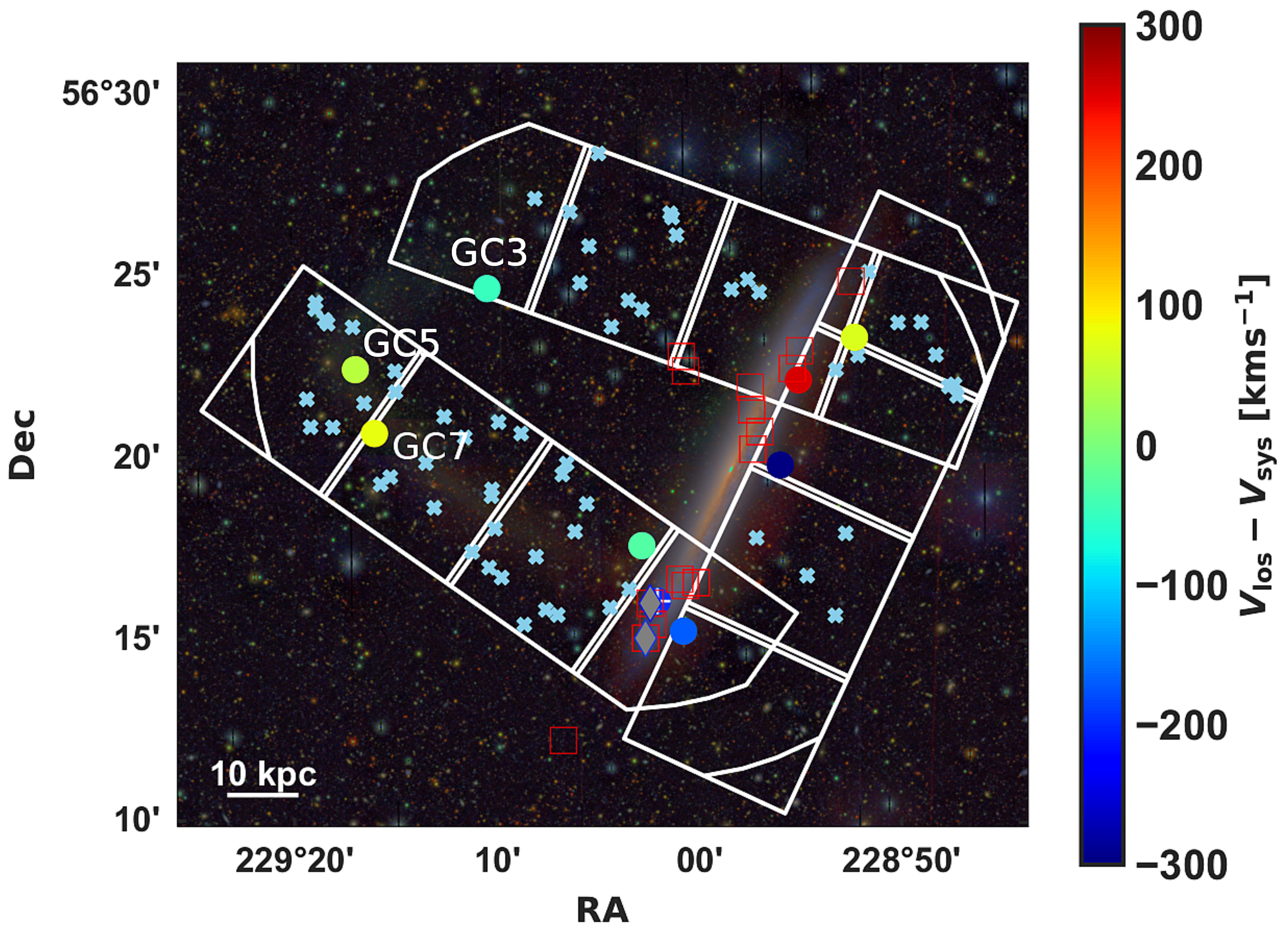}\hspace{0.01\textwidth}\\
	\caption{\label{fig:deimos_fig} The three observed DEIMOS masks overlaid on our \textit{gri} combined image. The skyblue crosses which dominate the three masks are the spectroscopically confirmed foreground stars. The filled circles, which have been color-coded according to their line-of-sight velocities, are the $8$ spectroscopically confirmed GCs. The stream GCs have been labeled (GC3, GC5 and GC7) while the rejected GC candidate based on our color selection criteria (see Section~\ref{color_selection}) is marked with a white cross. The filled gray diamonds and the open red squares are compact, likely young star cluster candidates with and without line-of-sight velocities, respectively. These young star cluster candidates are mostly spatially distributed along the outskirts of the galaxy disk.}
\end{figure*}
\begin{figure}
    \includegraphics[width=0.48\textwidth]{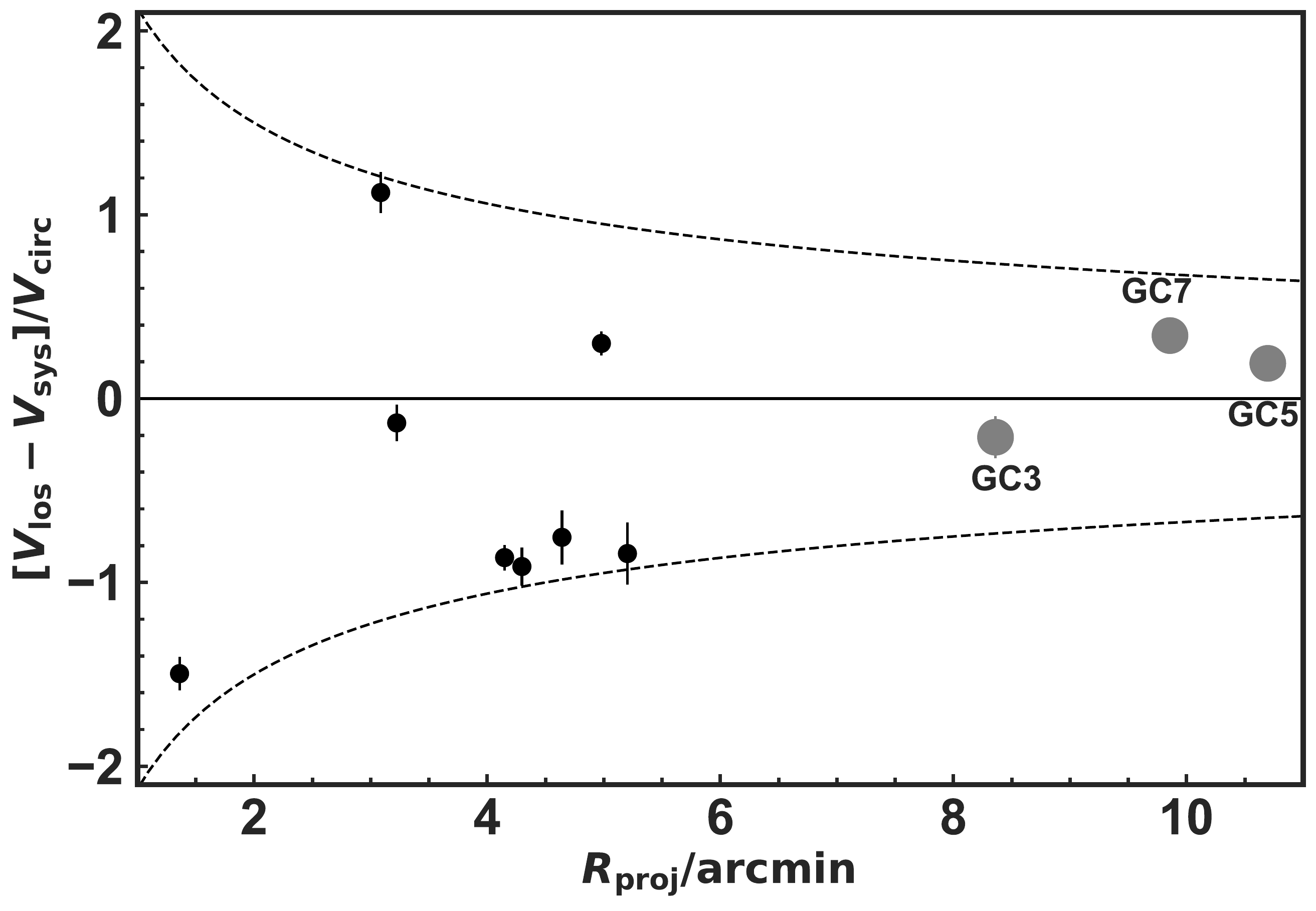}\hspace{0.01\textwidth}\\
	\caption{\label{fig:phase_space} Projected phase-space distribution of the $11$ point-sources belonging to NGC~5907. The gray filled circles are point-sources spatially associated with the stellar stream (GC3, GC5, and GC7) while the black filled circles are mostly near the galaxy disk. The dashed curves show the velocity limits consistent with the circular velocity of NGC~5907 assuming a logarithmic potential. The distribution of the stream GCs in phase-space is ``chevron-like'', consistent with expectations for cold kinematic substructures.}
\end{figure}

Assuming that all point-sources with measured $\vlos$ $<300~\kms$ are foreground stars, we identify $11$ point-sources belonging to NGC~5907 ($V_{\rm sys} =  667~\kms$) and/or the stream, and $64$ objects as Milky Way stars. These 11 point-sources have $300 \leq \vlos(\kms) \leq 920$ and $21.6 \leq i \leq 23.1$ with S/N per $\AA$ varying from $5-7$, but they are not all \textit{bona fide} GCs as we show in Section~\ref{YCs}. The foreground stars are relatively brighter, having $19.1 \leq i \leq 23.5$. One of the objects which we have classified as a star has $\vlos \sim 298~\kms$, very close to the limit we used to separate foreground stars from objects belonging to NGC~5907. However, it has both $g-i$ and $r-i$ colors outside the range we have used to define GCs and we therefore classify it as a foreground star. All other objects fall securely within the limits defined above. In addition, we also confirm $7$ background galaxies with redshifts ($z$) $\geq 0.01$ in the direction of NGC~5907. We show the projected phase-space distribution of the point-sources belonging to NGC~5907 in Figure~\ref{fig:phase_space} while we summarize all the point-sources with line-of-sight velocity measurements in Table~\ref{tab:complete_tab}. All spectroscopically confirmed foreground stars and background galaxies have been manually removed from our final GC catalog.

Out of the $11$ spectroscopically confirmed point-sources associated with NGC~5907 (Section~\ref{spec_GCs}), only $8$ appear to be GCs. Two, seen in projection near the galaxy disk, have very blue colors ($g - i \leq -0.1$) and are probably young star clusters (see Section~\ref{YCs}). The remaining point-source, which we exclude from further analysis, could be a \textit{bona fide} GC but it has a very red  color ($g - i$ $\sim1.6$) probably due to internal dust reddening and lies outside our GC  color--color selection region. Since the \textit{HST} study imaged the disk only, we are unable to spectroscopically confirm any of the \textit{HST}-detected GC candidates. It is particularly noteworthy that none of the 11 spectroscopically confirmed point-sources lie along the fainter E1 and E2 loops.

\section{RESULTS}
\label{partII}

\begin{table*}
\begin{tabular}{@{}l c c c c c c c c c c c c}
\hline
ID & RA & Dec & $g$  & $\Delta g$  & $r$    &  $\Delta r$    & $i$ & $\Delta i$ & $g-i$ & $r-i$ & $V$ & $\Delta V$ \\
     & [Degree] & [Degree]  & [mag] & [mag] & [mag]  &  [mag] & [mag] & [mag] &	[mag] & [mag] & [km s$^{-1}$]  & [km s$^{-1}$] 	                  \\
      (1)  & (2)      &  (3)            & (4)            & (5)     & (6)    &  (7)         &  (8)      &       (9)           &       (10)      & (11)  & (12) & (13)  \\
\hline
  NGC5907\_GC1  &   $228.9333 $ &  $56.330723$  &  $23.828$  &  $0.015$  &  $23.294$  & $0.015$  &  $22.837$  &  $0.015$  &  $0.99$ &  $0.46$ &  $328$  &  $20$ \\      
  NGC5907\_GC2  &   $229.01321$ &  $56.25453 $  &  $23.794$  &  $0.013$  &  $23.167$  & $0.011$  &  $22.898$  &  $0.014$  &  $0.9 $ &  $0.27$ &  $496$  &  $33$ \\      
  NGC5907\_GC3  &   $229.17627$ &  $56.411514$  &  $23.126$  &  $0.008$  &  $22.526$  & $0.007$  &  $22.345$  &  $0.009$  &  $0.78$ &  $0.18$ &  $619$  &  $26$ \\      
  NGC5907\_GC4  &   $229.04755$ &  $56.293774$  &  $23.195$  &  $0.009$  &  $22.676$  & $0.008$  &  $22.455$  &  $0.01 $  &  $0.74$ &  $0.22$ &  $637$  &  $22$ \\      
  NGC5907\_GC5  &   $229.28502$ &  $56.374107$  &  $22.799$  &  $0.007$  &  $22.139$  & $0.005$  &  $21.965$  &  $0.007$  &  $0.83$ &  $0.17$ &  $710$  &  $14$ \\    
  NGC5907\_GC6  &   $228.8717 $ &  $56.389336$  &  $22.432$  &  $0.005$  &  $21.902$  & $0.004$  &  $21.634$  &  $0.005$  &  $0.8 $ &  $0.27$ &  $735$  &  $14$ \\      
  NGC5907\_GC7  &   $229.26894$ &  $56.344845$  &  $22.421$  &  $0.005$  &  $21.851$  & $0.004$  &  $21.621$  &  $0.006$  &  $0.8 $ &  $0.23$ &  $745$  &  $11$ \\      
  NGC5907\_GC8  &   $228.91808$ &  $56.36979 $  &  $23.779$  &  $0.014$  &  $23.162$  & $0.012$  &  $22.845$  &  $0.015$  &  $0.93$ &  $0.32$ &  $921$  &  $25$ \\
\hline
  NGC5907\_YC1  &   $229.0406 $ &  $56.26739 $  &  $22.488$  &  $0.006$  &  $22.431$  & $0.007$  &  $22.571$  &  $0.012$  &  $-0.08$ & $-0.14$ &  $460$ &  $23$ \\    
  NGC5907\_YC2  &   $229.0445 $ &  $56.251354$  &  $22.785$  &  $0.007$  &  $22.878$  & $0.011$  &  $23.038$  &  $0.017$  &  $-0.25$ & $-0.16$ &  $476$ &  $38$ \\    
  NGC5907\_YC3  &   $228.93903$ &  $56.202152$  &  $23.483 $ &  $0.011$  &  $23.391$  & $0.013$  &  $23.44 $  &  $0.021$  &  $0.04$  & $-0.05$ &  $-$  &  $-$ \\   
  NGC5907\_YC4  &   $229.04419$ &  $56.20619 $  &  $ 23.512$ &  $0.01$   &  $23.332$  & $0.012$  &  $23.344$  &  $0.019$  &  $0.17$  & $-0.01$ &  $-$  &  $-$ \\  
  NGC5907\_YC5  &   $228.85585$ &  $56.23231 $  &  $23.598 $ &  $0.012$  &  $23.468$  & $0.014$  &  $23.431$  &  $0.023$  &  $0.17$  & $0.04 $ &  $-$  &  $-$ \\
$\cdots$  & $\cdots$  &	$\cdots$ &	$\cdots$ &	$\cdots$   &$\cdots$ &	$\cdots$ &	$\cdots$ &	$\cdots$ &	$\cdots$ & $\cdots$  &	$\cdots$ & $\cdots$\\
\hline
  NGC5907\_Star1 & $229.187898$ &  $56.290734$  &  $21.225$  &  $0.002$  &  $19.835$  & $0.001$  &  $19.079$  &  $0.001$ &  $2.15$  & $0.76$  &  $-61 $ &  $7 $ \\    
  NGC5907\_Star2 & $229.251732$ &  $56.363901$  &  $22.093$  &  $0.004$  &  $20.548$  & $0.001$  &  $19.162$  &  $0.001$ &  $2.93$  & $1.39$  &  $-63 $ &  $7 $ \\    
  NGC5907\_Star3 & $228.78685 $ &  $56.362426$  &  $19.671$  &  $0.001$  &  $19.302$  & $0.001$  &  $19.186$  &  $0.001$ &  $0.48$  & $0.12$  &  $-262$ &  $15$ \\    
  NGC5907\_Star4 & $228.788904$ &  $56.367609$  &  $19.994$  &  $0.001$  &  $19.434$  & $0.001$  &  $19.243$  &  $0.001$ &  $0.75$  & $0.19$  &  $-130$ &  $13$ \\    
  NGC5907\_Star5 & $229.171581$ &  $56.319447$  &  $21.573$  &  $0.002$  &  $20.198$  & $0.001$  &  $19.362$  &  $0.001$ &  $2.21$  & $0.84$  &  $16  $ &  $5 $ \\   
$\cdots$  & $\cdots$  &	$\cdots$ &	$\cdots$ &	$\cdots$   &$\cdots$ &	$\cdots$ &	$\cdots$ &	$\cdots$ &	$\cdots$ & $\cdots$  &	$\cdots$ & $\cdots$ \\
\hline
  NGC5907\_Gal1 &  $228.941315$ &  $56.25708 $  &  $23.267$  &  $0.009$  &  $22.48 $  & $0.007$  &  $22.253$  &  $0.009$  &  $1.01$  & $0.23$ &  $z=0.012$ & $-$ \\    
  NGC5907\_Gal2 &  $229.360868$ &  $56.343084$  &  $22.654$  &  $0.007$  &  $21.033$  & $0.002$  &  $20.489$  &  $0.002$  &  $2.16$  & $0.54$ &  $0.009$ & $-$ \\    
  NGC5907\_Gal3 &  $229.151877$ &  $56.443619$  &  $22.269$  &  $0.005$  &  $20.695$  & $0.002$  &  $20.153$  &  $0.002$  &  $2.12$  & $0.54$ &  $0.017$ & $-$ \\    
  NGC5907\_Gal4 &  $228.823128$ &  $56.348   $  &  $22.932$  &  $0.008$  &  $21.578$  & $0.004$  &  $21.069$  &  $0.004$  &  $1.86$  & $0.51$ &  $0.011$ & $-$ \\    
  NGC5907\_Gal5 &  $228.879945$ &  $56.327465$  &  $23.178$  &  $0.01 $  &  $22.774$  & $0.01 $  &  $22.505$  &  $0.012$  &  $0.67$  & $0.27$ &  $0.025$ & $-$ \\  
$\cdots$  & $\cdots$  &	$\cdots$ &	$\cdots$ &	$\cdots$   &$\cdots$ &	$\cdots$ &	$\cdots$ &	$\cdots$ &	$\cdots$ & $\cdots$  &	$\cdots$ & $\cdots$ \\
\hline   
\hline
\end{tabular}
\caption{Summary of globular clusters, young star clusters, stars and galaxies in the field of NGC~5907. \textit{Notes} Column (1): Object identifier, written as galaxy name 
and object type (globular clusters - GC, young star clusters - YC, stars and galaxies). Columns (2) and (3): Object position (RA and Dec, respectively) in degrees (J2000).
Columns (4)-(9): Subaru ${gri}$ photometry and corresponding uncertainties. The photometry has been corrected for Galactic extinction.  Columns (10) and (11): $(g-i)$ and $(r-i)$ colors. Columns (12) and (13): Measured line-of-sight velocities and uncertainties, respectively. We quote redshifts for background galaxies. The full version of this table is available in the online version but here we show all our spectroscopically confirmed GCs.}
\label{tab:complete_tab} 
\end{table*}

\subsection{Total number of Globular Clusters in NGC~5907}
\label{tot_G}

In order to obtain an estimate of the total number of GCs in NGC~5907, we need to account for the undetected GCs in the galaxy central region where the galaxy-light subtracted background varies anisotropically and also estimate the contaminants that might remain in our GC candidate catalog.

\subsubsection{Detection incompleteness in the galaxy central region}
\label{spatial_incomplete}

Within the region common to the \textit{HST} study, our GC detection is a factor of $\sim2$ incomplete relative to the \citet{KP_1999} sample. This incompleteness in detection implies that our GC catalog would be biased against the red GCs which are expected to dominate the galaxy central regions. We follow the method introduced by \citet{KP_1999} to estimate the number of undetected GCs within the region of maximum obscuration where we detect only $19$ GC candidates. If we assume that GCs in NGC~5907 and the Milky Way have similar spatial distributions, we estimate that $75$ GCs should be observed in this central region. We arrive at this estimate after projecting this region on the Milky Way GC system at the distance and orientation of NGC~5907 and also applying our magnitude incompleteness correction.

\begin{figure}
    \includegraphics[width=0.48\textwidth]{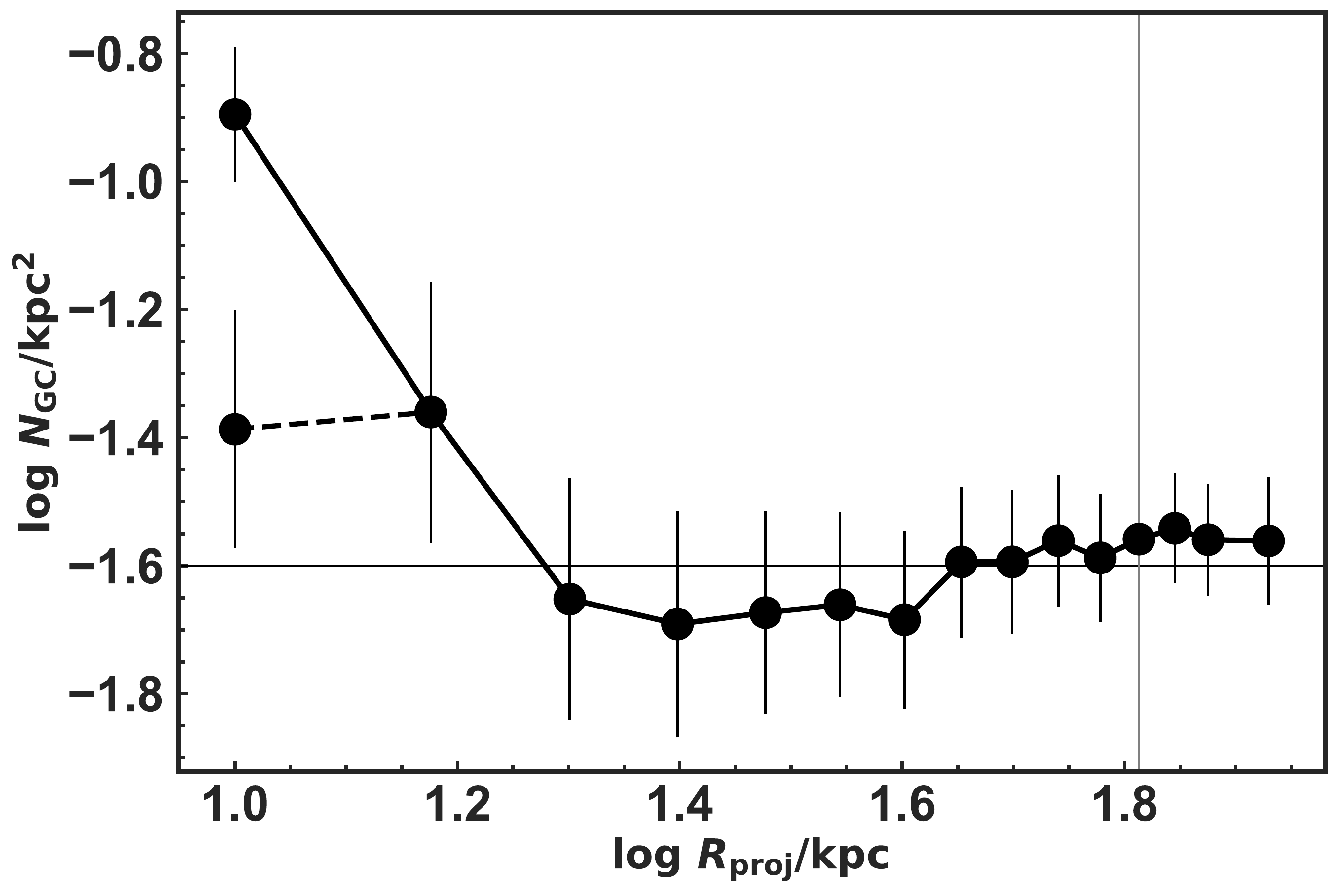}\hspace{0.01\textwidth}\\
	\caption{\label{fig:Nden} Radial number density profile of GC candidates in NGC~5907. The dashed and solid lines for the innermost datapoint show number density before and after correcting for undetected point sources in the galaxy central regions. The estimated background level is $\sim0.03$ objects per square kpc and is shown by the horizontal line. The vertical line corresponds to the radial extent of the NE stellar stream. Poisson errors are shown.
	}
\end{figure}

\begin{figure}
    \includegraphics[width=0.48\textwidth]{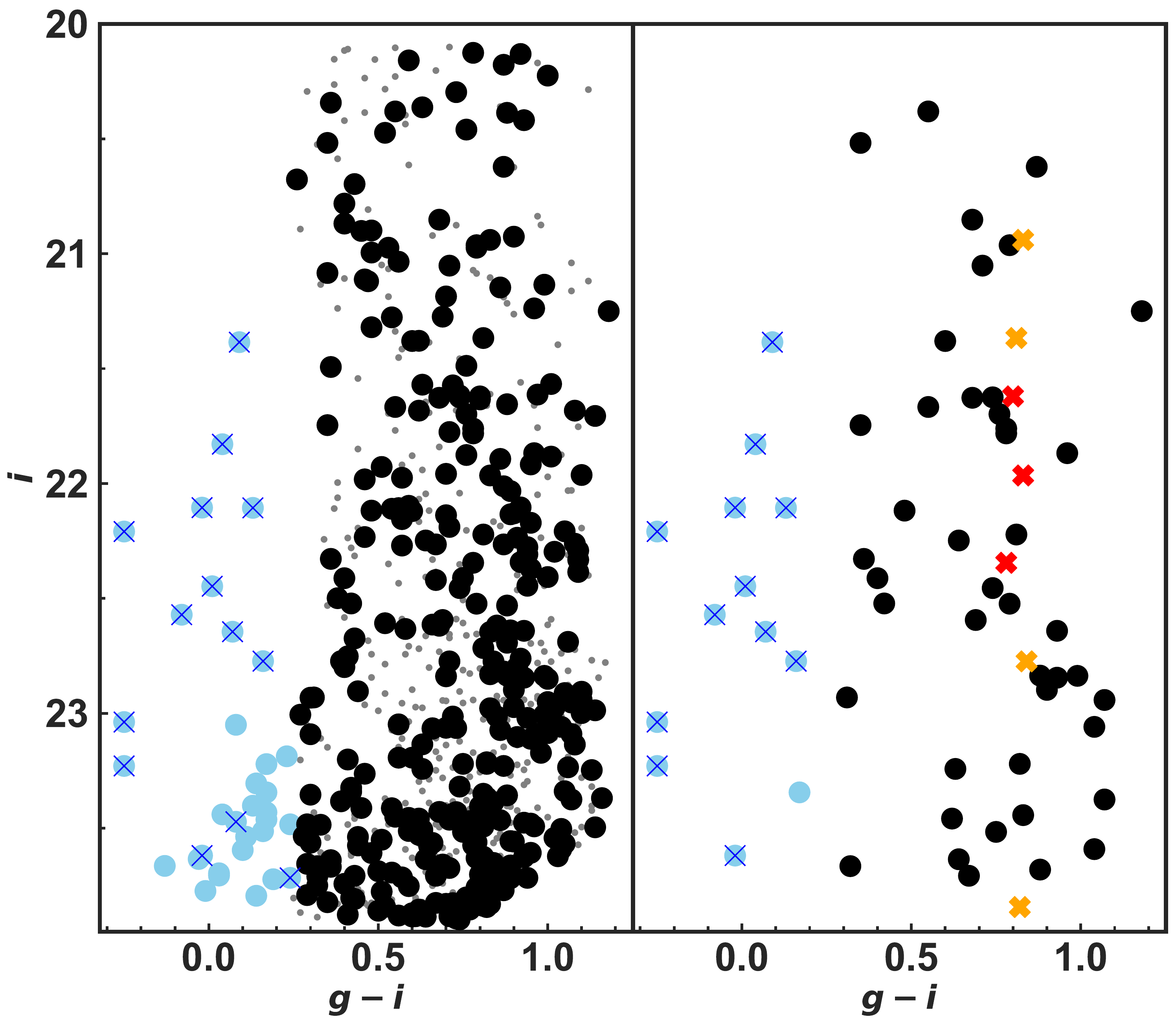}\hspace{0.01\textwidth}\\
	\caption{\label{fig:CMD} Color-magnitude diagrams of point-sources detected around NGC~5907. In both panels, globular cluster candidates defined within our magnitude, color, and spatial limits are shown as large black circles, while point-sources beyond $65$~kpc from the galaxy center are shown as grey dots. Young star cluster candidates are shown as skyblue circles, while those spatially associated with the galaxy disk are marked with blue crosses. The \textit{left panel} shows all the various selected point-sources while the \textit{right panel} shows the 45 GCs and the 13 young star cluster candidates in the central $\sim20$~kpc region. The   spectroscopically-confirmed globular clusters spatially associated with the stellar stream are shown by red crosses. The orange crosses are GC candidates on the stellar stream with no spectroscopic confirmation.}
\end{figure}

\subsubsection{Foreground and Background contamination}
\label{contamination}
We leverage the wide-field nature of our imaging to determine the contamination within our GC candidate catalog. Results from Section~\ref{spec_GCs}, where we found a low return rate of point-sources belonging to NGC~5907 (i.e. $11$ out of a total of $83$), already suggest a large fraction of contaminants in the field of NGC~5907. This is in agreement with previous results from \citet{Harris_1988} where they reported a  non-detection of GCs from their ground-based observations of NGG~5907. 

Figure~\ref{fig:Nden} shows the radial number density profile of all the GC candidates (including the contaminants). The number density in the flattened region of the profile (beyond the spatial extent of the NE stellar stream where contaminants are expected to dominate) corresponds to $\sim0.7$ objects per arcmin$^2$. Scaling this background region to match our entire field-of-view, we expect a total of $636$ contaminants (foreground MW stars and background galaxies) in our field. Using the Besan\c{c}on model \citep{Robin_2003} of the Galaxy, $480$ foreground stars are expected in the direction of NGC~5907 within our magnitude range and field of view. These contaminant estimates correspond to a number density of $0.52$ stars arcmin$^{-2}$ and $0.17$ background galaxies arcmin$^{-2}$.

The contaminant to GC ratio is consistent with the return rate from our spectroposcopy and gives a background-corrected, total GC candidate count ($N_{\rm GC}$) of $154\pm44$ after correcting for the undetected GCs and accounting for magnitude incompleteness. This value is a little lower than the $N_{\rm GC}$ estimated by \citet{KP_1999} but consistent within the uncertainty and corresponds to a specific globular cluster frequency of $S_{\rm N}$ of $0.73\pm0.21$.

\section{Young star clusters in NGC~5907}
\label{YCs}

An examination of the spatial distribution of the point-sources from Section~\ref{completeness} reveals an over-density at the intersection of the most prominent (NE) stellar loop with the galaxy disk. Visual inspection of these point-sources show that they are similar to GC candidates and two of them have been spectroscopically confirmed as members of NGC~5907, with line-of-sight velocities similar to the GCs. However they have very blue colors ($g-i < 0.3, r-i < 0.1$) and occupy the same region in color--color space as the young star clusters (YC) identified in M31 \citep[e.g.][]{Peacock_2009, Chen_2016}. We identify $34$ such point-sources as YC candidates, with $14$ of them spatially aligned along the outskirt of the galaxy disk as shown in  Figure~\ref{fig:deimos_fig}. Figure~\ref{fig:CMD} shows color-magnitude diagrams of our point-sources highlighting the young star cluster candidates. We note that similar to M31, the young clusters are systematically fainter than the globular clusters, suggesting that they may be less massive \citep{Caldwell_2009, Peacock_2009}. 

It is not obvious whether or not these YCs are part of a structure similar to the well-known ``ring of fire'' identified in M31 \citep[e.g.,][]{Brinks_1984}. There is however evidence that the two YC candidates with radial velocity measurements (see Figure~\ref{fig:deimos_fig}) are consistent with the overall disk  rotation. Their line-of-sight velocities (460 and 476~$\kms$) are consistent with the rotational velocity ($480\ \kms$) of the approaching arm from the HI velocity map \citep{Sancisi_1976}. Unfortunately, we do not have YCs spatially coincident with the receding arm. 

In their study of the stellar stream in NGC~5387, \citet{Beaton_2014} found a very-blue, young stellar overdensity at the intersection of the stream with the galaxy disk. They interpreted their result as evidence for either an induced star formation in the disk outskirts due to the satellite accretion event or starburst in the stream progenitor. \citet{Laine_2016}, in a stellar population study of the stellar stream in NGC~5907, reported no such stellar over-density nor did they find any blue star-forming regions. 
However it is evident from their figures 3 and 4 that they did not probe out to the region where we have observed these YCs.

\section{Stream Globular Clusters in NGC~5907}
\begin{figure}
   \includegraphics[width=0.48\textwidth]{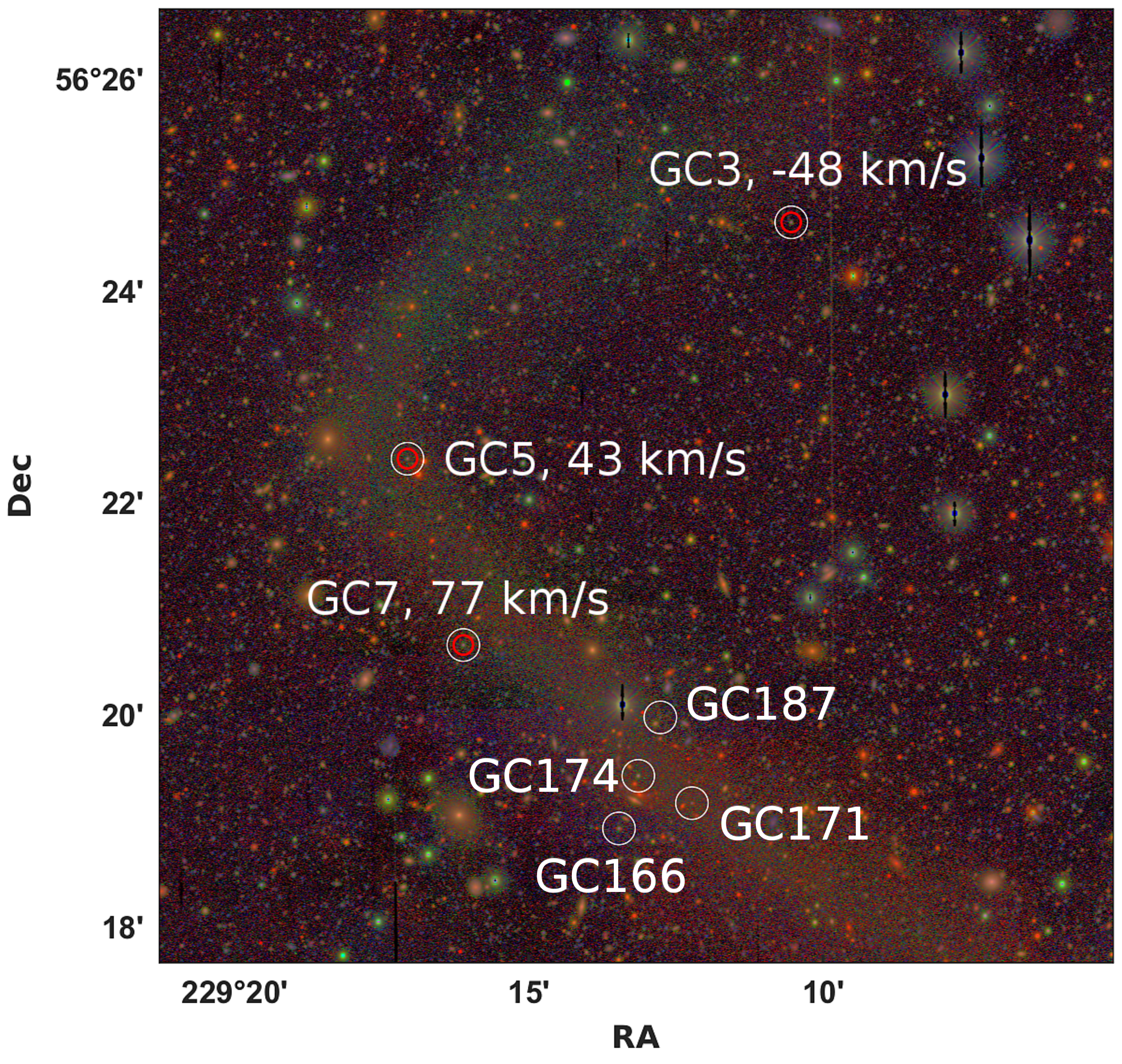}\hspace{0.01\textwidth}\\
	\caption{\label{fig:stream_vel} Globular clusters superimposed on the NE stellar loop. Red circles indicate spectroscopically confirmed GCs, and we show their line-of-sight velocities relative to NGC~5907. The white circles are the likely stream GC candidates with no velocity measurement.} 
\end{figure}

In this Section, we focus our attention on the $3$ spectroscopically confirmed GCs that are seen in projection along the most prominent NE stellar loop (i.e., GC3, GC5 and GC7). Figure~\ref{fig:phase_space} already shows that these stream GCs have the ``chevron-shaped'' distribution in phase-space that is usually associated with kinematic substructures \citep[e.g.][]{Romanowsky_2011}. These GCs have $g-i$ colors varying from $0.78$ to $0.83$ which corresponds to a mean [Fe/H] =  $-1.15\pm0.1$. We inferred this metallicity from a linear fit to the transformed $g-i$ colors and [Fe/H] of the Milky Way GC sample bearing in mind the recently reported variations in the GC color--[Fe/H] relation from galaxy to galaxy \citep{Villaume_2019}. Our inferred GC mean metallicity implies that the stellar mass of the progenitor galaxy is $\sim10^{8}~\Msun$, similar to the Sagittarius dwarf galaxy, using the stream GC metallicity--stellar mass relation from the E-MOSAICS simulations of \citet{Hughes_2018}. The stacked spectra shown in Figure~\ref{fig:stacked_spec} has a S/N per $\AA$ of $\sim12$ with strong Calcium triplet features consistent with typically old, metal-poor stellar populations. Assuming that GCs associated with the stellar streams all have similar colors, we find $4$ other plausible stream GC candidates (all along the NE loop) in our final catalog (see Table~\ref{tab:complete_tab}). These four objects, which lack spectroscopic confirmation, are highlighted in  Figure~\ref{fig:CMD} and are shown superimposed on the NE stellar loop in Figure~\ref{fig:stream_vel} and should be the subject of a follow-up spectroscopic study. 

\section{Summary and Conclusions}
\label{partIII}

In this work, we have used wide-field imaging and multi-object spectroscopy to investigate the GCs around the spiral galaxy NGC~5907 and its associated stellar stream from a disrupted satellite. We detect GC candidates out to $\sim65$~kpc from the galaxy center, with most of the spectroscopically confirmed candidates spatially distributed within $\sim20$~kpc of the galaxy disk. We estimate that NGC~5907 has a total of $154$ GCs and a specific frequency of $\sim0.7$. We identify GC candidates along the well--known NE stellar loop but spectroscopically find none along the fainter E1 and E2 loops. We also spectroscopically confirm five GCs spatially associated with the galaxy disk, and identify a new population of very blue, young star clusters mostly in the outskirts of galaxy disk, of which we spectroscopically confirm two. We provide an online catalog of all the detected sources in this work.

A deeper spectroscopic study of the GC candidates on the fainter loops as well as those photometrically identified on the NE loop is required to unambiguously resolve the nature and origin of the spectacular stellar stream in NGC~5907. This will allow for a straightforward comparison with the recent predictions from the E-MOSAICS cosmological simulations \citep{Hughes_2018} and various N-body models. Likewise, detailed stellar population study of the young star cluster candidates identified in this work is needed.

\section{Note Added in Proof}
After this paper was accepted, \citet{mller_2019} presented a deep imaging analysis of NGC~5907 in which they also found no evidence of the E1 and E2 loops.

\section*{Acknowledgements}

We thank the referee, David Mart{\'\i}nez-Delgado, for the thoughtful reading of the manuscript and for the valuable feedback.
We thank Kathryn Johnston for useful discussions. We thank Zach Jennings, Kristin Woodley, Asher Wasserman, Sabine Bellstedt, and Viraj Pandya for help with observations. DAF thanks the ARC for financial support via DP160101608. AJR was supported by National Science Foundation grant AST-1616710 and as a Research Corporation for Science Advancement Cottrell Scholar. JPB acknowledges financial support from AST-1616598 and AST-1518294. This paper was based in part on data collected at the Subaru Telescope, which is operated by the National Astronomical Observatory of Japan. Some of the data presented herein were obtained at the W. M. Keck Observatory, operated as a scientific partnership among the California Institute of Technology, the University of California and the National Aeronautics and Space Administration, and made possible by the generous financial support of the W. M. Keck Foundation. We wish to recognize and acknowledge the very significant cultural role and reverence that the summit of Mauna Kea has always had within the indigenous Hawaiian community. We are most fortunate to have the opportunity to conduct observations from this mountain.

\bibliographystyle{mnras}
\bibliography{n5907}

\appendix

\bsp
\label{lastpage}
\end{document}